\documentclass[a4paper,superscriptaddress,preprintnumbers,amsmath,amssymb,prb,floatfix,footinbib]{article}
\usepackage[scaled=2]{helvet}
\usepackage[T1]{fontenc}
\usepackage[latin9]{inputenc}
\usepackage{color}
\usepackage{textcomp}
\usepackage{amsmath}
\usepackage{amssymb}
\usepackage{graphicx}
\usepackage{setspace}
\usepackage{esint}
\usepackage[numbers]{natbib}
\makeatletter

\pdfpageheight\paperheight
\pdfpagewidth\paperwidth

\providecommand{\tabularnewline}{\\}

\newcommand{\lyxaddress}[1]{
\par {\raggedright #1
\vspace{1.4em}
\noindent\par}
}

\usepackage[scaled=2]{helvet}\usepackage{amstext}

\makeatletter
\@ifundefined{textcolor}{}{\definecolor{BLACK}{gray}{0}\definecolor{WHITE}{gray}{1}\definecolor{RED}{rgb}{1,0,0}\definecolor{GREEN}{rgb}{0,1,0}\definecolor{BLUE}{rgb}{0,0,1}\definecolor{CYAN}{cmyk}{1,0,0,0}\definecolor{MAGENTA}{cmyk}{0,1,0,0}\definecolor{YELLOW}{cmyk}{0,0,1,0}}

\usepackage{dcolumn}
\usepackage{bm}
\usepackage{times}
\usepackage{hyperref}\hypersetup{colorlinks=true,linkcolor=red,citecolor=blue}
\makeatother
\usepackage[english]{babel}

\makeatother

\usepackage{babel}

\makeatother

\begin{document}

\title{Lattice instabilities in bulk EuTiO$_{3}$}

\author{D. Bessas$^{1,2,\dagger}$, K. Z. Rushchanskii$^{3}$, M. Kachlik$^{4}$,
S. Disch$^{1,5}$, O. Gourdon$^{6,7}$,\\ J. Bednarcik$^{8}$, K.
Maca$^{4}$, I. Sergueev$^{1,8,}$, S. Kamba$^{9}$, M. Ležai\'{c}$^{3}$
and R. P. Hermann$^{1,2}$ \date{}}

\maketitle

\lyxaddress{$^{1}$Jülich Centre for Neutron Science JCNS and Peter Grünberg
Institut PGI, JARA-FIT, Forschungszentrum Jülich GmbH, D-52425 Jülich,
Germany}

\lyxaddress{$^{2}$Faculté des Sciences, Université de Liège, B-4000 Liège, Belgium}

\lyxaddress{$^{\dagger}$Present address: European Synchrotron Radiation Facility,
F-38043, Grenoble, France}

\lyxaddress{$^{3}$Peter Grünberg Institut, Quanten-Theorie der Materialien,
Forschungszentrum Jülich and JARA, D-52425 Jülich, Germany}

\lyxaddress{$^{4}$CEITEC Brno University of Technology, 61600, Brno, Czech Republic}

\lyxaddress{$^{5}$Institut Laue-Langevin, F-38042, Grenoble, France}

\lyxaddress{$^{6}$Neutron Scattering Science Division, Oak Ridge National Laboratory,
TN 37831, Oak Ridge, United States}

\lyxaddress{$^{7}$Jülich Centre for Neutron Science JCNS, Oak Ridge National
Laboratory, TN 37831, Oak Ridge, United States}

\lyxaddress{$^{8}$Deutsches Electronen-Synchrotron, D-22607 Hamburg, Germany}

\lyxaddress{$^{9}$Institute of Physics ASCR, 18221, Prague, Czech Republic}
\begin{abstract}
The phase purity and the lattice dynamics in bulk EuTiO$_{3}$ were
investigated both microscopically, using X-ray and neutron diffraction,
$^{151}$Eu-Mössbauer spectroscopy, and $^{151}$Eu nuclear inelastic
scattering, and macroscopically using calorimetry, resonant ultrasound
spectroscopy, and magnetometry. Furthermore, our investigations were
corroborated by $ab\textrm{ }initio$ theoretical studies. The perovskite
symmetry, $Pm\bar{3}m$, is unstable at the \emph{$M$-} and \emph{$R$-}
points of the Brillouin zone. The lattice instabilities are lifted
when the structure relaxes in one of the symetries: $I4/mcm$, \emph{$Imma$},
$R\bar{3}c$ with relative relaxation energy around$-25\textrm{ meV}$.
\textcolor{black}{Intimate phase analysis confirmed phase purity of
our ceramics.} A prominent peak in the Eu specific density of phonon
states at $11.5\textrm{ meV}$ can be modeled in all candidate symmetries.
A stiffening on heating around room temperature is indicative of a
phase transition similar to the one observed in SrTiO$_{3}$, however,
\textcolor{black}{although previous studies reported the structural
phase transition to tetragonal $I4/mcm$ phase our detailed sample
purity analysis and thorough structural studies using complementary
techniques did not confirm a direct phase transition.} Instead, in
the same temperature range, Eu delocalization is observed which might
explain the lattice dynamical instabilities.

\end{abstract}

\section{Introduction}

Perovskites exhibit cubic symmetry at high temperature, with space
group $Pm\bar{3}m$ and a large flexibility of site occupancy by a
broad range of elements on the $A$ and $B$ sites \citep{Li2004}
and flexibility in oxygen stoichiometry on the $X$ site \citep{Zhou2005},
indicated in the general chemical formula \emph{$ABX_{3\pm\mbox{\ensuremath{\delta}}}$}.
The oxygen in the perovskite unit cell form interconnected octahedra,
the rotation of which are potentially responsible for distortions
away from the cubic symmetry and structural transitions \citep{Glazer1972}.
Ba, Sr, and, to a lesser extent, Eu perovskite titanates are interesting
owing to their ferroelectric properties and their potential applications
in information technology. $\textrm{EuTi\ensuremath{\textrm{O}_{3}}}$
is an\textcolor{red}{{} }\textcolor{black}{incipient ferroelectric}
perovskite, like $\textrm{SrTi\ensuremath{\textrm{O}_{3}}}$, with
however magnetic cations on the $A$-site and it undergoes a transition
to a G-type antiferromagnetic phase below $5.3\textrm{ K}$ \citep{McGuire1966,Katsufuji2001}.
Recently, ferroelectric instability in $\textrm{EuTi\ensuremath{\textrm{O}_{3}}}$
films \citep{Lee2010} under $1\%$ tensile stress was reported. However,
in bulk $\textrm{EuTi\ensuremath{\textrm{O}_{3}}}$ the cubic $Pm\bar{3}m$
structure was till recently supposed to be stable down to LHe temperature
\citep{Brous1953}. Only lately some hints of an antiferrodistortive
phase transition to a tetragonal $I4/mcm$ phase were observed, but
the critical temperature varied from $200$ to $282\textrm{ K}$ \citep{Bussmann2011,Allieta2011,Kohler2012,Kim2012}.
These results were consistently checked by Goian $et\textrm{ }al$.
\citep{Goian2012} and attributed to the sample quality.

Inelastic X-ray scattering on tiny single crystals was only very recently
reported \citep{Ellis2012} and shows that the antiferrodistortive
phase transition cannot be of order - disorder type, as was proposed
by Bussmann-Holder $et\textrm{ }al.$ \citep{Bussmann2011,Bettis2011},
but is instead displacive with a soft mode at the $R$-point of the
Brillouin zone. However, similar zone boundary phonon softening has
been identified without the observation of a structural phase transition
by Swainson $et\textrm{ }al.$ \cite{Swainson2009} $e.g.$ in relaxors. 

In this study we present a detailed investigation of the phase purity,
the crystallinity and the lattice dynamics in bulk polycrystalline
$\textrm{EuTi\ensuremath{\textrm{O}_{3}}}$ using both microscopic
and macroscopic measurements as well as theoretical calculations.
We demonstrate that a lattice instability associated with Eu anharmonic
displacements appears close to room temperature.

\section{Methods}

\subsection{Experimental Techniques}

Phase pure EuTiO$_{3}$ polycrystalline samples were prepared using
a stoichiometric ratio of precursors (Eu$_{2}$O$_{3}$ 99.99\% and
Ti$_{2}$O$_{3}$ 99.9\%). The mixture was homogenised in a planetary
ball mill, cold isostatically pressed at $300\textrm{ MPa}$, and
sintered in pure hydrogen at $1400\textrm{ \textdegree C}$ for 2
hours. The processing details are given elsewhere \citep{Kachlik2012}.
The sintered pellets had a relative density of $89\%$. All measurements
were performed on ceramic pieces or powder taken from the same pellet.

\begin{figure}
\centering{}

\includegraphics{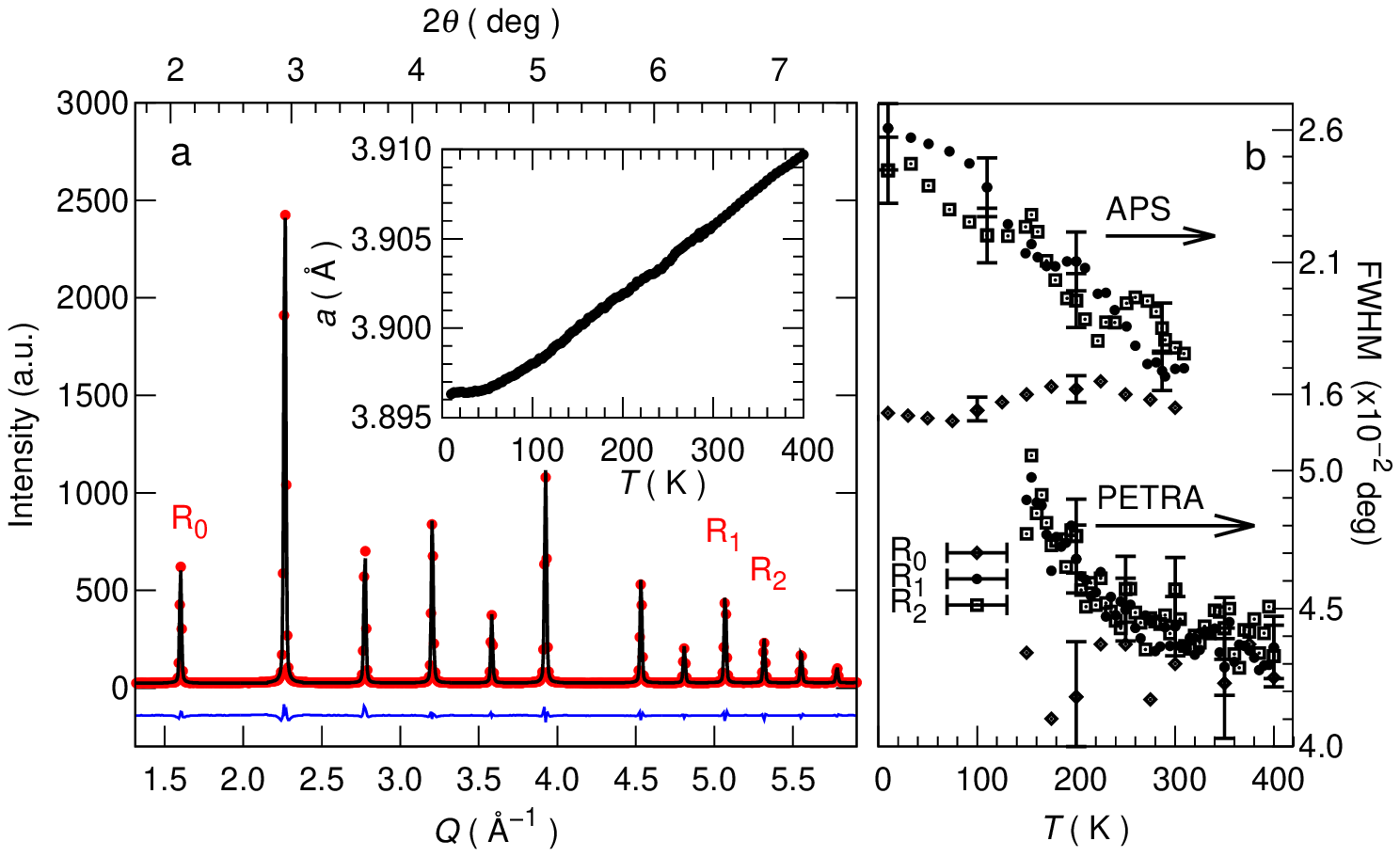}

\includegraphics{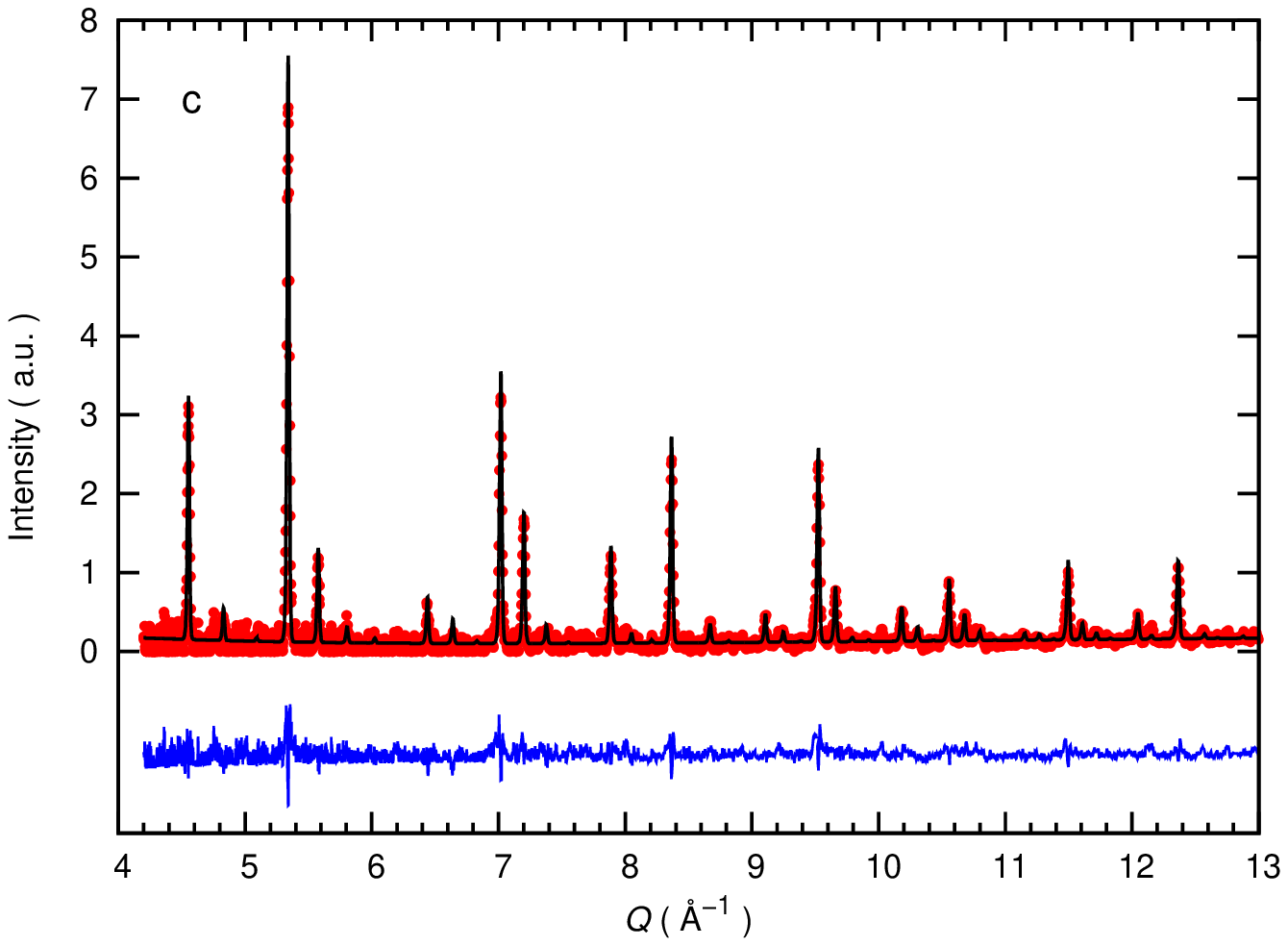}

\caption{\label{fig1} Rietveld refinement (black line) of a typical EuTiO$_{3}$
diffractogram (red points) and the corresponding refinement residuals
(blue line) using: (a) synchrotron radiation at $300\textrm{ K}$
and (c) neutrons at $180\textrm{ K}$. (b) The FWHM of certain reflections
(R$_{0}$, R$_{1}$ and R$_{2}$) \textcolor{black}{measured with
two different instrumental resolutions (see text).} Inset to (a) shows
the temperature dependent lattice parameter (pointsize defines errorbars)
in cubic symmetry. }
\end{figure}

The crystallographic phase purity and potential structural transitions
were checked by temperature dependent powder diffraction between $10$
and $300\textrm{ K}$ using high-energy synchrotron radiation, wavelength
$0.142013\textrm{ \AA}$, at station 6-ID-D/APS.\textcolor{black}{{}
In order to extend the temperature range up to $400\textrm{ K}$ similar
measurements were carried out with $0.20727\textrm{ \AA}$ wavelength
on the same sample at station P02.1/PETRAIII. The overall precision
including sample size and detector pixel size was estimated using
standard samples to $\Delta d/d\thicksim5.0\textrm{ x }10^{-3}$ for
measurements carried out at 6-ID-D and $\Delta d/d\thicksim1.0\textrm{ x }10^{-2}$
for measurements carried out at P02.1.} The atomic behaviour in $\textrm{EuTi\ensuremath{\textrm{O}_{3}}}$
was further studied by neutron diffraction. Europium is a strong neutron
absorber \citep{Ross1949}, thus a thin homogeneous powder layer was
prepared using $700\textrm{ mg}$ of $\textrm{EuTi\ensuremath{\textrm{O}_{3}}}$
and placed between thin vanadium foils ($0.02\textrm{ x }10\textrm{ x }30\textrm{ mm}^{3}$).
Neutron scattering data were collected between $150$ and $350\textrm{ K}$
using the time-of-flight, T.O.F., instrument POWGEN \citep{Huq2011},
with precision $\Delta d/d\thicksim1.5\textrm{ x }10^{-3}$ at $d=1\textrm{ Å}$,
and between $10$ and $300\textrm{ K}$ using the T.O.F. instrument
NOMAD \citep{Neuefeind2012} at the Spallation Neutron Source. The
neutron pair distribution function, PDF, analysis was carried out
on the data obtain at NOMAD by Fourier transformation of the total
scattering function \citep{Egami2003}. 

The phase purity was further investigated using $^{151}$Eu-Mössbauer
spectroscopy between $90$ and $325\textrm{ K}$ on fine powder of
$\textrm{EuTi\ensuremath{\textrm{O}_{3}}}$, $35\textrm{ mg/c\ensuremath{\textrm{m}^{2}}}$,
mixed with BN, using a calibrated spectrometer. 

Heat capacity measurements were recorded in the Quantum Design cryostat
(QD-PPMS) utilising the built-in calorimeter. Measurements of both
the addenda and sample were performed at the same temperatures between
$10$ and $340\textrm{ K}$ with a $0.5\textrm{ K}$ point density
in the region of interest. Every data point was measured three times
and an average value was extracted. 

The macroscopic lattice dynamics was probed by Resonant Ultrasound
Spectroscopy (RUS) \citep{Migl1993}. Temperature dependent spectra
between $100$ kHz and a few MHz were recorded on a shaped sample
($2.5\textrm{ x }2\textrm{ x }1.5\textrm{ mm}^{3}$) using an in-house
spectrometer made with cylindrical Y-cut LiNbO$_{3}$ $0.3\textrm{ mm}$
thick transducers (Ø $1.5\textrm{ mm}$) inside a QD-PPMS. 

Magnetic characterizations was carried out in a Cryogenics Ltd. measurement
system using AC susceptibility measurements below $30\textrm{ K}$
( $f=20.4\textrm{ Hz}$, $H_{\textrm{ac}}=10\textrm{ G}$ at $H_{\textrm{dc}}=0\textrm{ G}$)
as well as DC susceptibility measurements ( $H_{\textrm{dc}}^{\textrm{max}}=5\textrm{ kG}$,
at $T=4.9\textrm{ K}$).

To support our macroscopical lattice dynamical characterization microscopic
investigations based on nuclear inelastic scattering \citep{Alp2002},
NIS, of $^{151}\textrm{Eu}$ in $\textrm{EuTi\ensuremath{\textrm{O}_{3}}}$
were performed. Several spectra were recorded at $110$, $210$, $295$
and $360\textrm{ K}$ in 16-bunch mode at the nuclear resonance station
ID22N/ESRF \citep{Rueffer1996} using a nested monochromator \citep{Leupold1996}
providing $1.5\textrm{ meV}$ resolution.

\begin{figure}
\centering{}\includegraphics{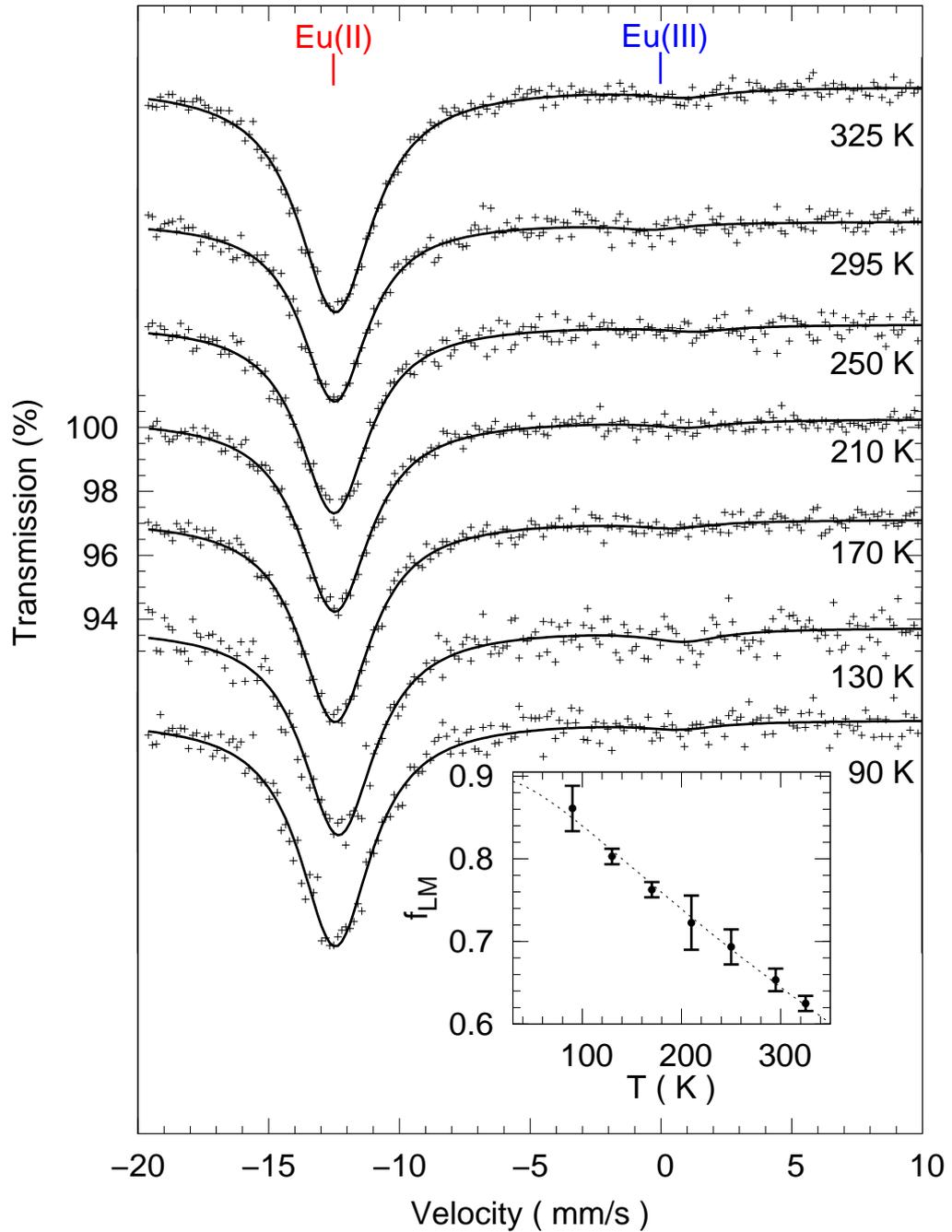}\\
 \caption{\label{fig2}Temperature dependent$^{151}$Eu-Mossbauer spectra measured
on $\textrm{EuTi\ensuremath{\textrm{O}_{3}}}$ powder (black points)
and the corresponding two component model (black line). The expected
isomer shift relative to EuF$_{3}$ is indicated with tics. The inset
shows the temperature dependent Lamb-Mössbauer factor, $f_{\textrm{LM}}$,
of the majority phase, Eu(II), extracted from the spectra and the
associated Debye model (black dashed line) .}
\end{figure}

\subsection{Theoretical Techniques}

The Eu specific density of phonon states, $i.e.$ the spectral distribution
of the Eu vibrational amplitude, for $I4/mcm$ , \emph{$Imma$} and
$R\bar{3}c$ structures were calculated using the PHON code \citep{Alfe2009}
by integrating the eigenvectors and eigenvalues of the corresponding
dynamical matrices over the Brillouin zone. We used a $30\textrm{ x }30\mbox{ x }30$
$q$-point mesh and the experimental FWHM for Gaussian convolution.
Details on the calculation of the dynamical matrices as well as the
crystallographic parameters of the used structures are given in Ref.
\cite{Rushchanskii2012}.

\section{Results}

\subsection{Microscopic Characterization}

A typical X-ray diffractogram recorded on polycrystalline EuTiO$_{3}$
is shown in Fig.\ref{fig1}a. The $Q$ range of our scattering data
reached approximately $13\textrm{ Å}^{-1}$ using neutrons. A typical
T.O.F. neutron diffractogram is given in Fig.\ref{fig1}c. All observed
reflections were identified in all possible symmetries. No peak splitting
characteristic of structural phase transitions was observed in our
temperature dependent measurements using both X-rays and neutrons.
In order to account for any resolution limited satellite reflection
the Full Width Half Maximum, FWHM, of all reflections was studied
using a Lorentzian profile. The extracted FWHM of selected reflections
between $10$ \textcolor{black}{and $400\textrm{ K}$}\textcolor{red}{{}
}(indicated in Fig.\ref{fig1}a by $\textrm{R}_{1}\equiv\left(\textrm{3 1 0}\right)$
in $Pm\bar{3}m$ or \{$\left(\textrm{1 3 4}\right)$,$\left(\textrm{1 2 8}\right)$\}
in $R\bar{3}c$ or \{$\left(\textrm{1 1 6}\right)$, $\left(\textrm{3 3 2}\right)$,
$\left(\textrm{4 2 0}\right)$\} in $I4/mcm$ or \{$\left(\textrm{0 6 4}\right)$,
$\left(\textrm{2 6 0}\right)$, $\left(\textrm{0 2 12}\right)$, $\left(\textrm{2 0 12}\right)$,
$\left(\textrm{6 2 0}\right)$, $\left(\textrm{6 0 4}\right)$\} in
$Imma$ and $\textrm{R}_{2}\equiv\left(\textrm{3 1 1}\right)$ in
$Pm\bar{3}m$ or \{$\left(\textrm{0 4 2}\right)$, $\left(\textrm{2 2 6}\right)$,
$\left(\textrm{0 2 10}\right)$\} in $R\bar{3}c$ or \{$\left(\textrm{4 2 2}\right)$,
$\left(\textrm{2 0 6}\right)$\} in $I4/mcm$ or \{$\left(\textrm{2 6 4}\right)$,
$\left(\textrm{2 2 12}\right)$, $\left(\textrm{6 2 4}\right)$\}
in $Imma$) are shown in to Fig.\ref{fig1}b. The FWHM of all the
examined reflections \textcolor{black}{which are supposed to split
broadens upon cooling below $\sim300\textrm{ K}$. This is not the
case for $\textrm{R}_{0}\equiv\left(\textrm{1 0 0}\right)$ in $Pm\bar{3}m$
or $\left(\textrm{0 1 2}\right)$ in $R\bar{3}c$ which shows temperature
independent FWHM. Although the observation of temperature dependent
FWHM in some particular reflections m}ight be indicating departure
from cubic symmetry,\textbf{ }new reflections were not observed in
our diffractogram. The diffraction data were further refined with
Fullprof \citep{Rodri1993} using the Rietveld method. A typical refinement,
with $R_{\textrm{wp}}=8.3\textrm{ \%}$, is shown in Fig.\ref{fig1}a.
The inset to Fig.\ref{fig1}a shows the extracted lattice parameter
in cubic symmetry $vs$ temperature. The extracted lattice parameter
is in excellent agreement with reference data \citep{Brous1953}.
Linear thermal expansion is observed between $100$ and $300\textrm{ K}$.
The calculated volume thermal expansion coefficient, $\alpha_{\textrm{V}}$,
after fitting the lattice parameter with a linear function normalised
to the lattice parameter at $300\textrm{ K}$ is $\alpha_{\textrm{V}}=9.9(1)\textrm{ x }10^{-6}\textrm{ K}^{-1}$.

\begin{figure}
\begin{centering}
\includegraphics{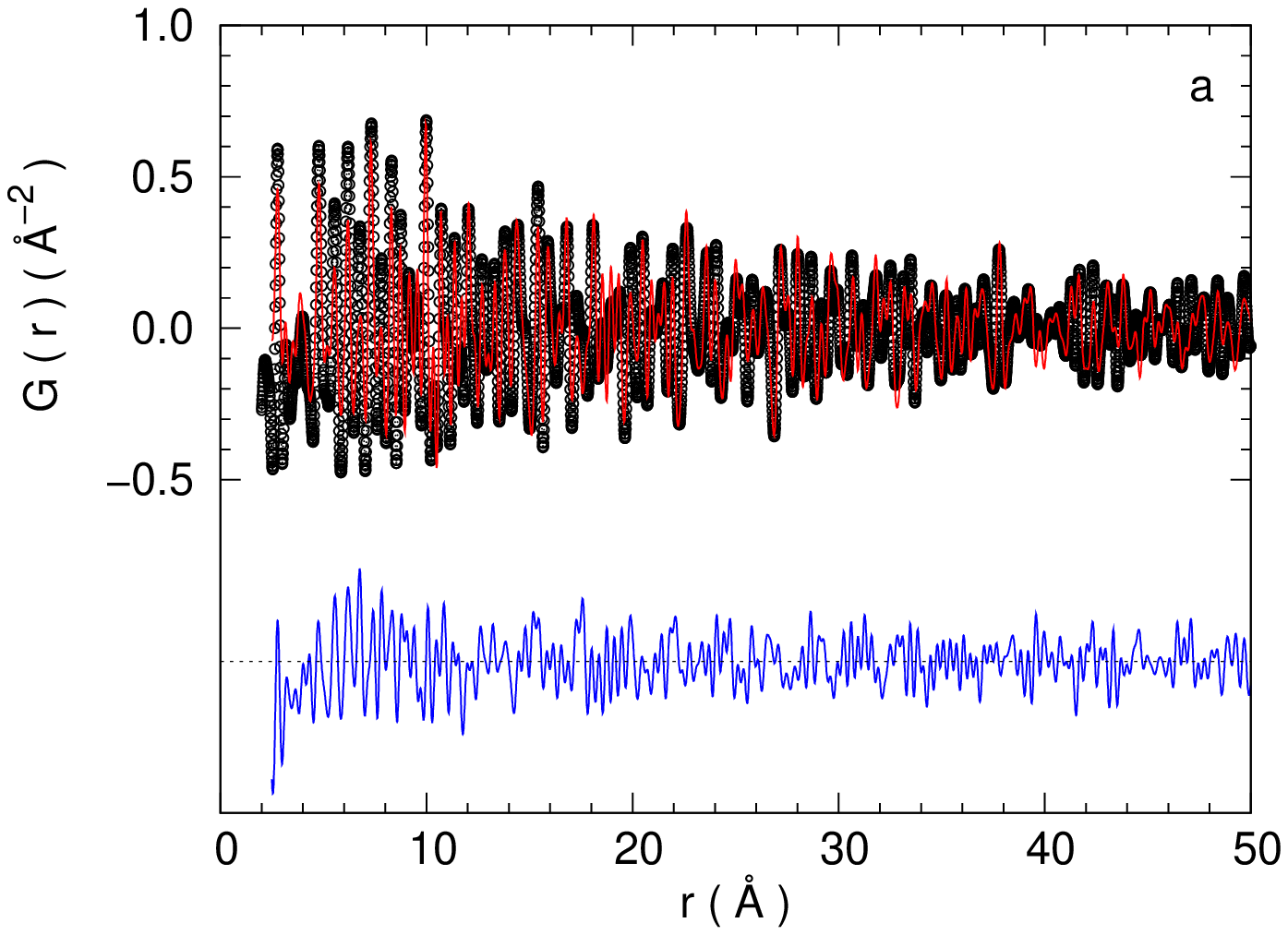}
\par\end{centering}

\begin{centering}
\includegraphics{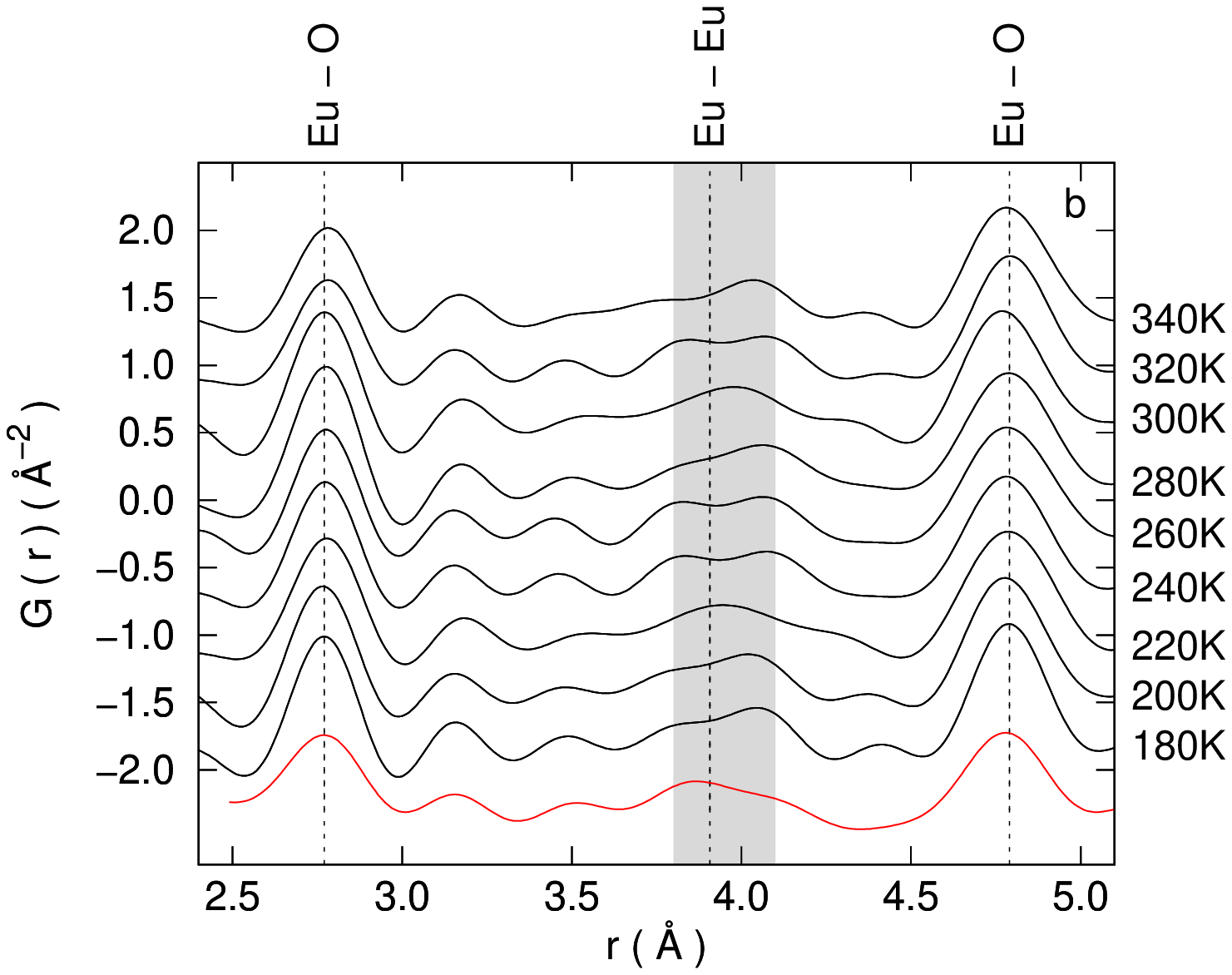}
\par\end{centering}

\centering{}\caption{\label{fig3-1}(a) Pair Distribution Function analysis (PDF) of neutron
scattering data, $Q=50\textrm{ Å}$, (black points), at $300\textrm{ K}$,
and the corresponding refinement (red line). The refinement residual
(blue line) is given in the same scale and represents a fair refinement.
(b) shows a close up in the temperature dependent PDF analysis (black
lines) and a typical PDF analysis refinement (red line) around the
first nearest neighbor distances regime, indicated with dashed lines.
Note that at a lattice parameter distance, $3.905\textrm{ Å}$ at
RT, Ti-Ti and O-O pair correlations coexist with Eu-Eu. }
\end{figure}

Minor deviations from linearity are observed between $200$ and $270\textrm{ K}$
which might support the claim of instabilities in this region. However,
no sign for a phase transition is observed.

The neutron diffraction data obtained at POWGEN were further refined
with JANA2006 \citep{JANA2006} using the Rietveld method resulting
in $R_{\textrm{wp}}=7.4\%$. The extracted lattice parameter in cubic
symmetry is in excellent agreement with the one extracted using X-rays.
In addition, the atomic displacement parameters, ADP, of Eu, Ti and
O at $180\textrm{ K}$ extracted in the harmonic approximation are
$21.0$$(8)$, $7.3(5)$ and $7.5(5)\textrm{ (}\textrm{x}10^{-3}\textrm{ Å}^{2})$
respectively and do not show any substantial irregularity $vs$ temperature.
Note that in isostructural SrTiO$_{3}$ the atomic displacements of
Sr and Ti in the antiferrodistortive phase are comparable \citep{Kopecky2012}
to that of Eu in EuTiO$_{3}$.

Typical Mössbauer spectra between $90$ and $325\textrm{ K}$ are
given in Fig.\ref{fig2}. The data were fitted with a two component
model, the second component contributing by less than $1(1)\textrm{ }\%$
to the total area. The extracted isomer shift for the major component
is $-12.45(5)\textrm{ mm/s}$ relative to $\textrm{Eu\ensuremath{\textrm{F}_{3}}}$,
an isomer shift indicative of Eu(II). Thus, the sample contained purely
divalent Eu. The upper limit of Eu(III) which might escape detection
is $1\%$. A Debye temperature of $295(5)\textrm{ K}$, was calculated
within the Debye approximation \citep{Herber1984} from the temperature
dependent Lamb - Mössbauer factor, $f_{\textrm{LM}}$. In contrast
to the prediction by Bussmann-Holder $et\textrm{ }al.$ \cite{Bussmann2011}
no line splitting or broadening was found in our temperature dependent
Mössbauer spectra. 

\begin{figure}
\centering{}\includegraphics{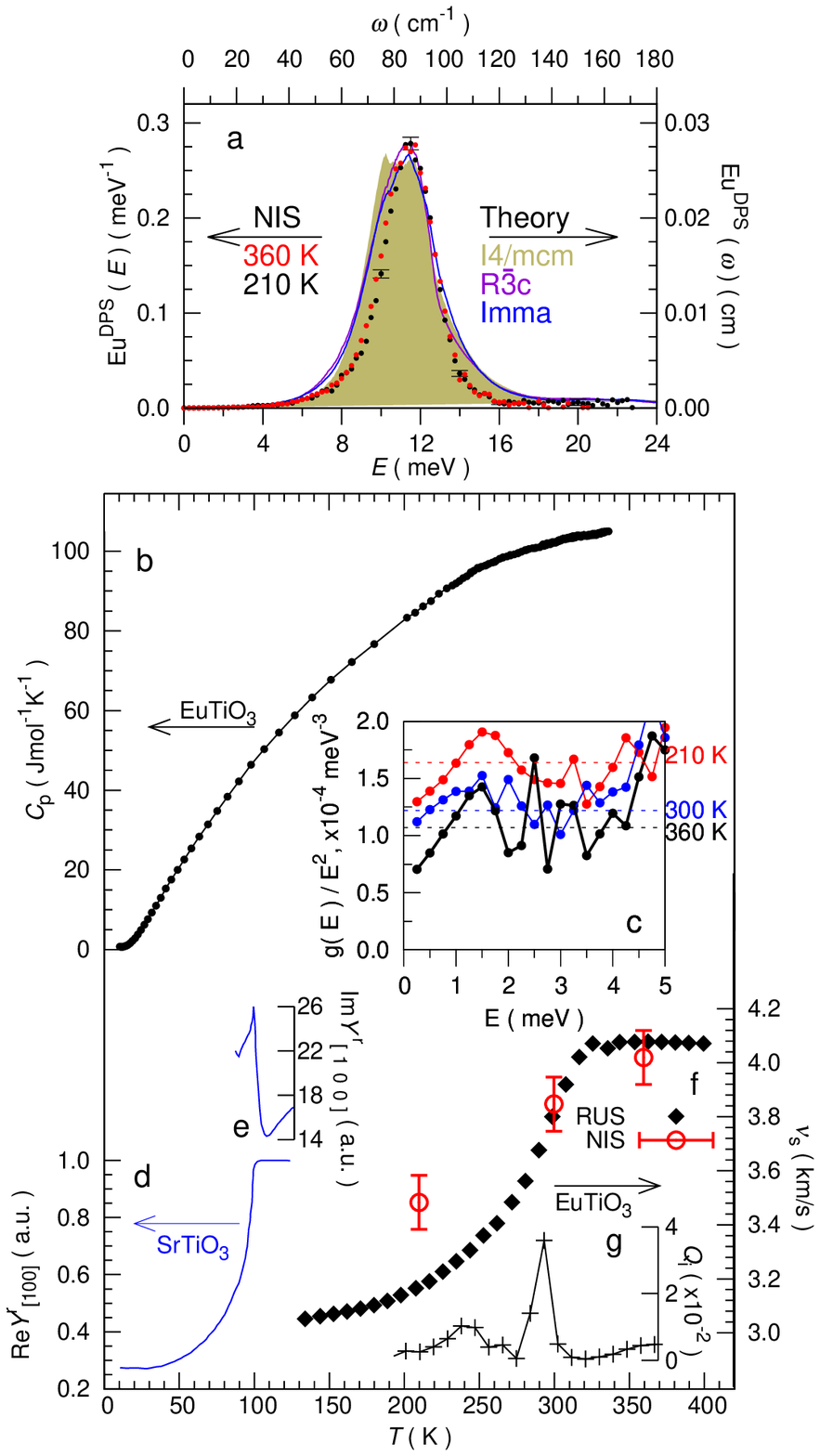}
\caption{\label{fig4_a} (a) The Eu specific density of phonon states in EuTiO$_{3}$
at $210\textrm{ K}$ (black tics) and $360\textrm{ K}$ (red line-tics)
measured using NIS, typical errorbar is given, and our theoretical
spectra for different structures\citep{Rushchanskii2012}. \label{fig4_b}
(b) Heat capacity data between $10$ and $340\textrm{ K}$ measured
on EuTiO$_{3}$ using calorimetry (pointsize defines errorbar), line
between points is guide to the eye. \label{fig4_c} (c) The first
$5\textrm{ meV}$ of the reduced $^{151}$Eu projected density of
phonon states, $g(E)/E^{2}$, and the related Debye levels (dashed
lines). \label{fig4_d}(d) The real part, $\textrm{Re}Y_{[1\textrm{ }0\textrm{ }0]}^{\textrm{r}}$,
and the imaginary part, (e) \label{fig4_e}, $\textrm{Im}Y_{[1\textrm{ }0\textrm{ }0]}^{\textrm{r}}$,
of the complex Young's modulus in SrTiO$_{3}$ obtained from Ref.
\citep{Kityk2000}. \label{fig4_f}(f) Speed of sound extracted from
RUS (black points), pointsize defines errorbar, and the corresponding
extracted from NIS (red circles) at $210$, $300$ and $360\textrm{ K}$.
\label{fig4_g} (g) The inverse quality factor of the resonance at
$590\textrm{ kHz}$ \textcolor{black}{between $200$ and $370\textrm{ K}$.} }
\end{figure}

Pair Distribution Function analysis (PDF) probes local disorder in
crystalline materials \citep{Qiu2005,Jeong2007}. In contrast to Rietveld
refinements, the diffuse scattering and other background contributions
are of crucial importance because a Fourier transformation is applied
to the total scattering function. In this case, the large neutron
absorption combined with the time-of-flight instrument prevent accurate
background subtraction and produces oscillations in the extracted
PDF thus in this study background contributions are treated phenomenologically.
The refinement was conducted using PDFgui \citep{Farrow2007} between
$2.5$ and $50\textrm{ Å}$, see Fig.\ref{fig3-1}a. Within the limited
precision of our extracted PDF no clear change with temperature in
the interatomic distances of oxygen with europium is observed, see
Fig.\ref{fig3-1}b. 

Several experimental methods for probing lattice dynamics exist. However,
access to the full Density of Phonon States (DPS), $g(E)$, is feasible
only by inelastic neutron \citep{Christensen2006} or X-ray scattering
\citep{Burkel2000}. In this work, nuclear resonance inelastic measurements
which requires the existence of Mössbauer active isotope and synchrotron
radiation were carried out. The raw spectra were treated using a modified
version \citep{DOS} of the program DOS \citep{Kohn2000}. The $^{151}$Eu-specific
density of phonon states \citep{151Eu}, DPS, was extracted between
$0$ and $24\textrm{ meV}$, see Fig.\ref{fig4_a}a. A single peak
around $11.5\textrm{ meV}$ is observed, in agreement with our first
principles calculations in all possible symmetries. No resolvable
change has been observed in the $^{151}$Eu DPS between $210$ and
$360\textrm{ K}$. The $^{151}$Eu mean force constant can be extracted
from our data using the expression $\left\langle F_{i}\right\rangle =M_{i}\intop_{0}^{\infty}g(E)E^{2}dE/\hbar^{2}$,
where $M_{i}$ is the mass of the resonant isotope. Between $110$
and $360\textrm{ K}$ the extracted $^{151}$Eu mean force constant
ranges from $78$ to $70\textrm{ N/m}$. From our NIS data the Eu
ADP, $\left\langle u^{2}\right\rangle $, were extracted using $\left\langle u^{2}\right\rangle =-lnf_{\textrm{LM}}/k^{2}$,
where $k$ is the wavenumber of the resonant photons. $\left\langle u^{2}\right\rangle $
are in fair agreement with these extracted from neutron diffraction
due to incoherent - coherent origin, respectively \citep{Sales2001}.
In the long wavelength limit, in this work assumed below $4\textrm{ meV}$,
the average speed of sound, $\nu$$_{s}$, can be extracted from the
DPS using: lim$_{E\rightarrow0}$$\frac{g(E)}{E^{2}}=\frac{M_{\textrm{i}}}{2\pi\hbar^{3}\rho v_{s}^{3}}$
\citep{Ashcroft1976} where $M_{\textrm{i}}$ is the isotopic mass
and $\rho$ is the mass density. A linear fit of the $g(E)/E^{2}$,
below $4\textrm{ meV}$, between $110$ and $360\textrm{ K}$ is given
in Fig. \ref{fig4_c}c. The extracted speed of sound is included in
the same figure.

\subsection{Macroscopic Characterization}

In order to verify claims of a \textit{striking} phase transition
observed in the heat capacity, $C_{\textrm{p}}$, of $\textrm{EuTi\ensuremath{\textrm{O}_{3}}}$
\citep{Bussmann2011,Petrovic2013} the same cryostat, Quantum Design,
was used. Special attention was taken on the thermal coupling between
the measuring platform and the sample \citep{QD2002}. Every data
point was measured three times and an average value has been extracted.
The averaged data are shown in Fig.\ref{fig4_a}b. The measured heat
capacity in $\textrm{EuTi\ensuremath{\textrm{O}_{3}}}$ reveals no
evidence of a structural phase transition in contrast with what has
been observed in Ref. \cite{Bussmann2011, Petrovic2013} and using
similar techniques in $\textrm{SrTi\ensuremath{\textrm{O}_{3}}}$
\citep{Gallardo2002}. Hence, we conclude that the observation reported
in Ref. \cite{Bussmann2011,Petrovic2013} is related either to sample
purity or inadequate background subtraction. 

The isotropic elastic tensor, $C_{11}$ and $C_{44}$, was extracted
from the spectrum of resonant ultrasound spectroscopy and the bulk,
$B_{295\textrm{ K}}=125\textrm{ GPa}$, and shear, $G_{295\textrm{ K}}=76\textrm{ GPa}$,
moduli were calculated. The extracted polycrystal shear wave speed,
$\nu$$_{\textrm{s}}$$,$ is shown in Fig.\ref{fig4_a}f. In Fig.\ref{fig4_a}g
the inverse quality factor, $Q_{\textrm{i}}$\citep{Que}, of a typical
mechanical resonance at $590\textrm{ kHz}$ is shown. The average
speed of sound calculated from the isotropic elastic tensor indicates
$25\%$ hardening at $300\textrm{ K}$ relative to $100\textrm{ K}$.
Similar behaviour was observed on a second sample from the same batch.

\begin{figure}
\centering{}\includegraphics{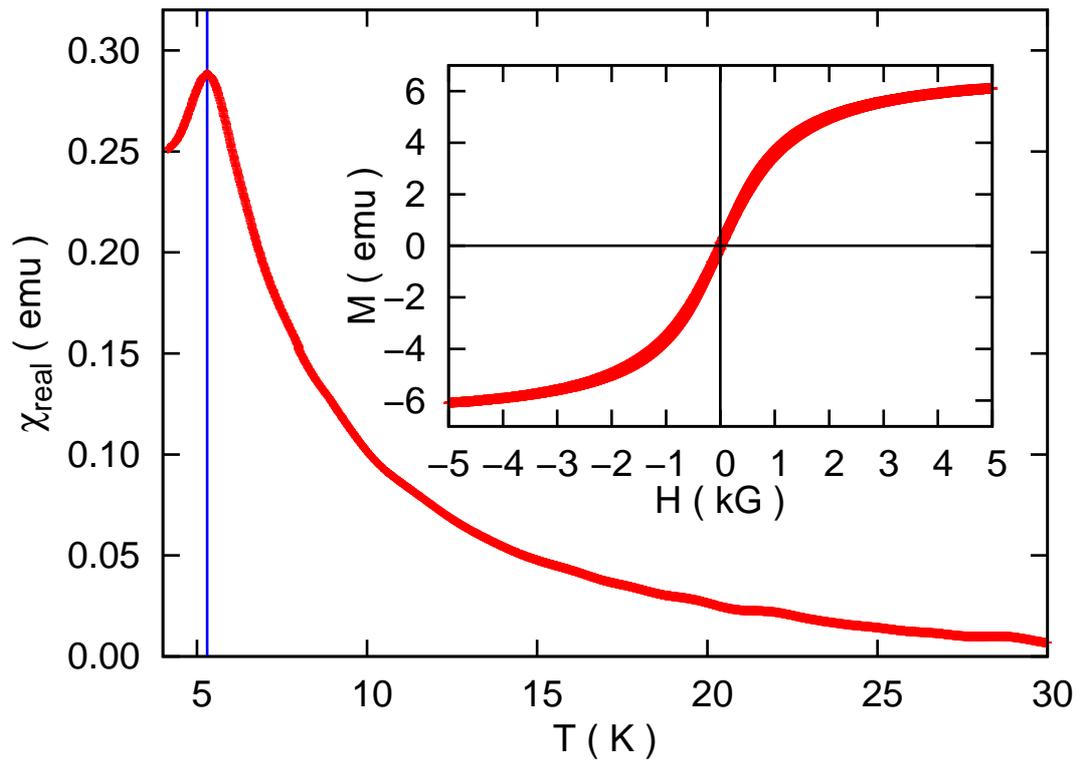}
 \caption{\label{fig5} Temperature dependence of the ac magnetic susceptibility
measured on cooling in sample in $\textrm{EuTi\ensuremath{\textrm{O}_{3}}}$
using an oscillating magnetic field with frequency $f=20.4\textrm{ Hz}$
amplitude $H_{\textrm{ac}}=10\textrm{ G}$ at $H_{\textrm{dc}}=0\textrm{ G}$.
Inset shows a $M-H$ curve measured on the same sample at $4.9\textrm{ K}$
with maximum applied magnetic field of $H_{\textrm{dc}}=50\textrm{ kG}$. }
\end{figure}

In our magnetic characterization measurements, see Fig. \ref{fig5},
a prominent peak at $T=5.2(1)\textrm{ K}$ appears in magnetic susceptibility.
No other magnetic transitions were identified below $30\textrm{ K}$.
Below the observed transition, at $4.9\textrm{ K}$, dc magnetization
was measured up to $\mu_{0}H_{\textrm{dc}}=5\textrm{ T}$ and neither
a hysteretical behavior nor a ferromagnetic contribution were observed.

\subsection{Theoretical Investigations}

Our first-principle calculations \citep{Rushchanskii2012} on $\textrm{EuTi\ensuremath{\textrm{O}_{3}}}$
show that the \emph{$Pm\bar{3}m$} symmetry is unstable at the \emph{$M$-}
and \emph{$R$-} points of the Brillouin zone with respect to the
rotation of the oxygen octahedra. The lattice instabilities are removed
when the structure relaxes in one of three symmetries: tetragonal
($I4/mcm$), orthorhombic (\emph{$Imma$}) or rhombohedral ($R\bar{3}c$)
all with relative relaxation energies between $-25$ and $-27\textrm{ meV}$
to the cubic symmetry. The energy difference between the distorted
structures is very small and allows in fact polymorphism. The Eu specific
density of phonon states in all calculated structures is in excellent
agreement with the measured one, see Fig.\ref{fig4_a}a. We note that
although in $Imma$ phase a Eu atomic displacement is allowed and
its calculated size \citep{Rushchanskii2012} is $0.012\textrm{ Å}$. 

\section{Discussion}

Although no obvious structural instability was identified by diffraction
or calorimetry, a prominent acoustical stiffening upon heating is
nevertheless observed between $100$ and $300\textrm{ K}$, a stiffening
which is coroborated by a peak in the quality factor, $Q_{\textrm{i}}$,
at $290\textrm{ K}$. \textcolor{black}{In the same temperature range
the diffraction peaks which are suppose to split upon phase transition
broaden on cooling, see Fig.\ref{fig1}. Above $310\textrm{ K}$ the
FWHM of the same reflections as well as the speed of sound are found
to be essentially temperature independent. }A similar behaviour was
observed in the real and imaginary part of Young's modulus around
$100\textrm{ K}$ in $\textrm{SrTi\ensuremath{\textrm{O}_{3}}}$ \citep{Kityk2000},
see Fig. \ref{fig4_d}d and Fig. \ref{fig4_e}e, however, in $\textrm{SrTi\ensuremath{\textrm{O}_{3}}}$
a well known structural transition around $100\textrm{ K}$ takes
place \citep{Lytle1964}. Based on ultrasonic experiments Rehwald
\cite{Rehwald1973} qualifies the observed phase transition in $\textrm{SrTi\ensuremath{\textrm{O}_{3}}}$
as a second order and not as order-disorder but as soft mode type.
In order to understand the nature of the accoustical stiffening in
$\textrm{EuTi\ensuremath{\textrm{O}_{3}}}$ which appears rather similar
to the one observed in $\textrm{SrTi\ensuremath{\textrm{O}_{3}}}$
detailed structural and phase purity analysis was carried out. 

Divalent Eu based titanium compounds, such as $\textrm{EuTi\ensuremath{\textrm{O}_{3}}}$,
have been investigated extensively by McGuire $et\textrm{ }al.$ \citep{McGuire1963}
owing to their remarkable variety in magnetic properties. Among them,
pyrochlore compound Eu$_{3}$Ti$_{2}$O$_{7}$ and Eu$_{2}$TiO$_{4}$
perovskite are impurity candidates in any EuTiO$_{3}$ sample. According
to literature\citep{Greedan1972}, both compounds show ferromagnetic
transitions around $8.7\pm0.3\textrm{ K}$ . However, in our magnetization
data, see Fig.\ref{fig5}, neither ferromagnetic transition appeared
around $9\textrm{ K}$ nor is a ferromagnetic contribution present
in the $M-H$ curve below the antifferomagnetic transition temperature.
As a result, within our instrumental resolution and with the combined
use of several characterization methods, neither Eu$_{3}$Ti$_{2}$O$_{7}$
nor Eu$_{2}$TiO$_{4}$ are present in our sample. Thus combining
the results of magnetic characterization with the Mössbauer spectroscopy
our sample properties are consistent with the reported antiferromagnetic
properties of $\textrm{EuTi\ensuremath{\textrm{O}_{3}}}$ \citep{Greedan1972}
and preclude other europium titanates \citep{Henderson2007,Syamala2008,McCarthy1969}. 

Mössbauer spectroscopy and nuclear inelastic spectroscopy are complementary
methods since they are based on the same principle. This is illustrated
once more by the fact that the NIS extracted $f_{\textrm{LM}}$ agrees
within $95\%$ with the one extracted by Mössbauer spectroscopy. From
the long wavelength limit, below $4\textrm{ meV}$, of the DPS the
speed of sound, $\nu$$_{s}$, was calculated. The DPS extracted speed
of sound indicates hardening of $\textrm{EuTi\ensuremath{\textrm{O}_{3}}}$
versus temperature, as observed also using RUS, see Fig. \ref{fig4_a}f.
Both microscopic and macroscopic measurements are in good agreement.
The $10\%$ deviation at $210\textrm{ K}$ is reproducible and can
be associated with isothermal speed of sound measured via scattering,
and adiabatic sound measured with RUS. As a result, the increase in
speed of sound upon heating is verified both by microscopic and macroscopic
techniques and is in contrast to the usual softening of elastic constants
upon heating \citep{Varshni1970} confirming lattice instability of
$\textrm{EuTi\ensuremath{\textrm{O}_{3}}}$ between $100$ and $300\textrm{ K}$.
The temperature behaviour of the adiabatic speed of sound is not consistent
with an order - disorder transition\citep{Rehwald1973}. 

Pair Distribution Function analysis (PDF) probes local disorder in
crystalline materials \citep{Qiu2005,Jeong2007}. The PDF can be derived
either from X-ray or neutron total scattering data with advantages
and disadvantages described extensively by Egami and Billinge \citep{Egami2003}.
In this study we carried out PDF analysis on our neutron data because
if there was any potential oxygen displacement it would be more visible
using neutrons due to the enhanced neutron cross section with respect
to X-rays. The main information extracted from PDF without further
modeling is the interatomic distances, see Fig.\ref{fig3-1}b. \textcolor{black}{Ti
has a negative coherent scattering length which results in negative
peaks for the $A$,$X$-Ti correlations, where $A$ and $X$ are the
perovskite sites.} Europium neutron absorption introduces anomalous
background in the total scattering as function of the scattering angle.
Hence, the Fourier transformation of the total scattering function
might introduce artifacts in the PDF which cannot be modeled.\textcolor{black}{{}
}Fig.\ref{fig3-1}a shows the room temperature PDF of EuTiO$_{3}$.
Although all interatomic distances were modelled sucessfully and the
extracted parameters (lattice parameters, atomic displacement parameters)
agree with those extracted from Rietveld refinement, the goodness
of fit is relatively poor. In case there was a pronounced phase transition,
the interatomic distance between Eu - Eu, Ti - Ti and O - O it would
change and a doublet instead of single peak is expected at the corresponding
interatomic distance. However, the highlighted part of Fig.\ref{fig3-1}b
does not show any consistent variation. The expected Eu - Eu correlation
peak, and similarly the Ti - Ti and O - O, is not observed probably
because the variations in PDF introduced by Eu absorption was not
modelled correctly.

\begin{figure}
\centering{}\includegraphics{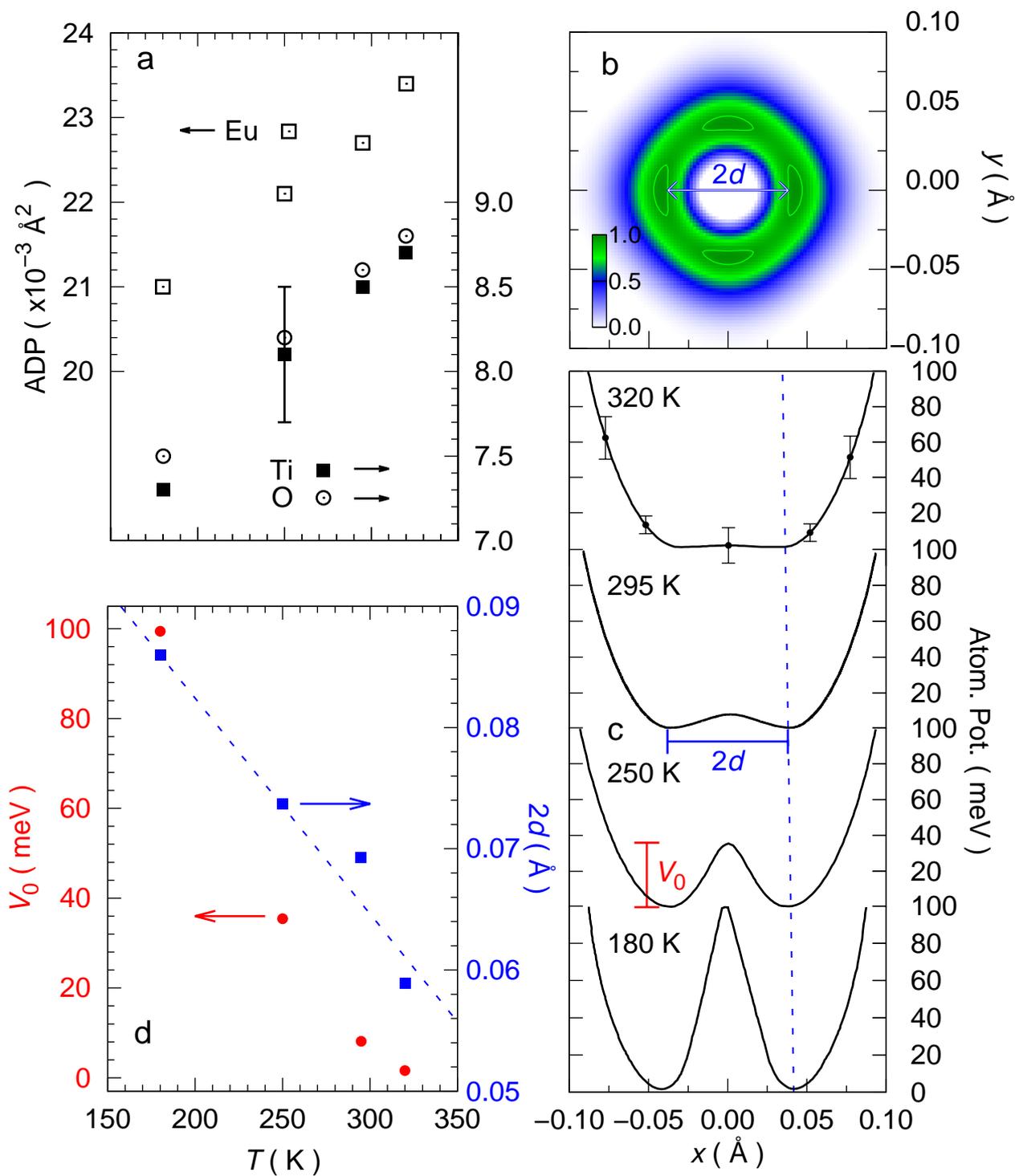}
\caption{\label{fig6} (a) The atomic displacement parameters, ADP, extracted
in the harmonic approximation from neutron diffraction, typical errorbar
is given. (b) Probability density function, p.d.f, distribution of
the Eu-atom in the $ab$ plane at 180 K. (c) The effective one-particle
potential of the Eu atom along the $a$ direction ($y$ = 0) at $320$,
$295$, $250$ and $180\textrm{ K}$ obtained from crystallographic
structure analysis and the typical errorbar. (d) The potential barrier,
$V_{0}$, and the displacement from equilibrium position, $2d$, (dashed
line is guide to the eye) extracted from the potential given in (c).}
\end{figure}

\begin{table}[h]
\caption{Temperature dependent anharmonic refined parameters $D_{\textrm{GC}}^{ijkl}(r)$
of Eu atom in $\textrm{EuTi\ensuremath{\textrm{O}_{3}}}$ refined
using the Gram-Charlier expansion }

\begin{tabular}{ccccc}
 & \multicolumn{4}{c}{Temperature ( K )}\tabularnewline
$i,j=1,2,3$ & $320$ & $295$ & $250$ & $180$\tabularnewline
\hline 
$D_{\textrm{GC}}^{iiii}(r)$  & $-426.0(1)$ & $-375.6(1)$ & $-329.8(1)$ & $-88.7(1)$\tabularnewline
$D_{\textrm{GC}}^{iijj}(r)$  & $-65.7(8)$ & $-67.5(6)$ & $-69.2(7)$ & $-35.1(4)$\tabularnewline
\hline 
\end{tabular}\label{Table} 
\end{table}

Although absolute values of ADP extracted using coherent techniques
such as diffraction might be affected by occupancies, relative changes
of the ADP extracted from such measurements are still valid. Although
all extracted ADP have smooth dependence on temperature the Eu ADP
is large as compared to Ti and O, see Fig. \ref{fig6}a. Hence, a
Fourier map study in the vicinity of Eu was carried out. The Gram-Charlier
expansion of anharmonic atomic displacement parameters is extensively
described in Ref. \cite{Kuhs1992} and has been followed in several
cases of perovskite structure \citep{Etschmann2001,Zhurova2000} with
 reported anharmonic atomic displacements. In this study, the Eu atomic
displacement parameters extracted from neutron diffraction assuming
cubic symmetry, $Pm\overline{3}m$ , was modeled using a Gram-Charlier
expansion of the probability density function, p.d.f., $p_{\textrm{Eu}}^{\textrm{GC}}$,
up to fourth-rank tensor\citep{ITfC1995} given in Eq. \ref{eq:S1}.

\begin{align}
p_{\textrm{Eu}}^{\textrm{GC}}(r) & =p_{\textrm{Eu}}^{\textrm{harm}}(r)[1+\frac{1}{4!}D_{\textrm{GC}}^{ijkl}(r)H_{ijkl}(r)]\label{eq:S1}
\end{align}
where $r$ is the atomic displacement vector relative to the equilibrium
position, $H_{ijkl}(r)$ is the Hermite polynomial of fourth order
and $D_{\textrm{GC}}^{ijkl}(r)$ are anharmonic refined parameters
(the third-order cumulants, $C_{\textrm{GC}}^{ijkl}(r)$ are zero,
based on the site symmetry). The use of a fourth order Gram-Charlier
expansion \citep{Kuhs1992} for the Eu ADP improves the refinement.
A reduction of the reliability factor from $5.03\%$ to $4.07\%$
just by adding a single parameter $D_{\textrm{GC}}^{1111}(r)$ has
been observed. The refined Eu anharmonic parameters $D_{\textrm{GC}}^{ijkl}(r)$
at $110$, $210$, $295$ and $360\textrm{ K}$ are given in the Table
\ref{Table}. Indeed, the Gram-Charlier tensor is following the symmetry
restriction and therefore for the Eu at $(0,0,0)$; $C_{\textrm{GC}}^{ijkl}(r)$
are zero and $D_{\textrm{GC}}^{ijkl}(r)$ are non zero only for $D_{\textrm{GC}}^{1111}(r)=D_{\textrm{GC}}^{2222}(r)=D_{\textrm{GC}}^{3333}(r)$
and $D_{\textrm{GC}}^{1122}(r)=D_{\textrm{GC}}^{1133}(r)=D_{\textrm{GC}}^{2233}(r)$.
To justify the existence of such anharmonic behavior on the Eu site
similar treatment has been attempted on the Ti and O sites. No major
refinement improvement has been observed and thus\textbf{ }we have
no evidence for strong anharmonic behavior on the Ti and O sites.
In Fig.\ref{fig6}b the europium p.d.f. after final refinement at
$180\textrm{ K}$ is illustrated. Eu exhibits off-centering in $\left[1\textrm{ }0\textrm{ }0\right]$
and $\left[0\textrm{ }1\textrm{ }0\right]$ directions (and equivalently
in the $\left[0\textrm{ }0\textrm{ }1\right]$ direction) with significant
residual probability density in the azimuthal direction. The effective
one-particle atomic potential, $V(r)$, is related to the p.d.f. by:

\begin{equation}
V(r)=-\textrm{k}_{\textrm{B}}T\textrm{ln}\left[p_{\textrm{Eu}}^{\textrm{GC}}(r)/p_{\textrm{Eu}}^{\textrm{GC}}(r_{0})\right]\label{eq:S1-1}
\end{equation}
where $\textrm{k}_{\textrm{B}}$ is the Boltzmann constant and $T$
is temperature. In Fig.\ref{fig6} a section of the Eu one-particle
potential extracted according to the Eq. \ref{eq:S1-1} is depicted
along the $\left[1\textrm{ }0\textrm{ }0\right]$ direction. These
sections reveal that Eu exhibits temperature dependent off-centering,
see Fig.\ref{fig6}d, with $d\sim0.04\textrm{ Å}$ at $180\textrm{ K}$.
In addition, the potential barrier along the azimuthal direction flattens
well before $295\textrm{ K}$,\textcolor{black}{{} whereas at $250\textrm{ K}$
a double well potential is already formed. Note that this is the same
temperature region in which specific diffraction peaks, see Fig. \ref{fig1},
start to broaden upon cooling. }Above $295\textrm{ K}$, the Eu p.d.f
forms a plateau. An analogous double-well potential for EuTiO$_{3}$
was suggested theoretically by Bettis $et\textrm{ }al.$ \citep{Bettis2011}
although for oxygen displacement. 

The plateau in the p.d.f. indicates increased Eu anharmonicity. In
order to quantify anharmonicity, we estimated using our macroscopic
measurements of $C_{\textrm{p}}$, $\alpha_{\textrm{V}}$, $B$ and
the Grüneisen rule \citep{Gopal1966} the Grüneisen parameter. Our
estimation of $\gamma$ at $290\textrm{ K}$ is, $1.3(1)$, in the
same range with typical metallic compound\textcolor{black}{s, }$\gamma\simeq2$\textcolor{black}{.
To elucidate the impact of the Grüneisen parameter on our measured
DPS we used the vibrational frequency definition of the Grüneisen
parameter, }$\gamma=-\frac{dlnE}{dlnV}$\textcolor{black}{, which
relates the change in phonon mode energy to the change in volume.
The estimated phonon mode energy shift, between $110$ and $360\textrm{ K}$
using our measured }$\frac{dV}{V}\sim0.0025$\textcolor{black}{{} and
our extracted average Grüneisen parameter of }$\gamma=1.3$\textcolor{black}{{}
results in }$\frac{\delta E}{E}\sim0.004$\textcolor{black}{. Therefore,
the prominent peak, around $11\textrm{ meV}$ will not shift due to
anharmonicity by more than $0.04\textrm{ meV}$. Such energy mode
shift is currently resolution limited.}

EuTiO$_{3}$ is an incipient ferroelectric, which means that its lattice
is close to a ferroelectric instability at low temperatures. Note
that theoretical calculations\citep{Rushchanskii2012} indicate that
the Eu displacement from its high-symmetry position is possible in
tilted structures, $e.g.$ $Imma$, however, this displacement is
static. Significant dynamical displacement of Eu was observed in cubic
$Pm\overline{3}m$ structure even for the low-energy optical mode
TO1, which eigenvector constitutes $37\%$ of Last mode and $60\%$
of the Slater, see Ref. \cite{Rushchanskii2012}. This observation
was used to design ferroelectirc $\textrm{(Eu,Ba)Ti}\textrm{O}_{3}$
ceramics with significant off-centering on the magnetic cation site
\citep{Rushchanski2010}. Also magnetoelectric coupling in EuTiO$_{3}$
is exceptionally high \citep{Shvartsman2010} due to strong contribution
of magnetic Eu cation in low energy polar mode. Note that in other
multifferroics non-magnetic ions are displaced, $e.g.$ Bi in BiFeO$_{3}$
or Y in YMnO$_{3}$, therefore the magnetoelectric coupling is usually
smaller in such materials. Microscopic origin of the mixing character
of the low-energy polar mode in cubic $\textrm{EuTi}\textrm{O}_{3}$
is the coupling of the 4f orbitals of Eu$^{2+}$ with the 3d states
of nonmagnetic Ti$^{4+}$\cite{Akamatsu2011}. The influence of this
coupling on the lattice instabilities of $\textrm{EuTi}\textrm{O}_{3}$
was recently studied by Birol and Fennie \cite{Birol2013}. It was
found that partial occupation of the d-states on Ti due to hybridization
drives $\textrm{EuTi}\textrm{O}_{3}$ away from ferroelectric instability.
This conclusion is compatible with results of Ref. \cite{Rushchanskii2012},
where it was shown that increasing volume of cubic $\textrm{EuTi}\textrm{O}_{3}$,
$i.e.$ decreasing f-d hybridization, turns the low-energy mode to
be unstable, with Slater type atomic displacement. Nevertheless, it
is still not clear what is the effect of the oxygen vacancy on the
structural stability of $\textrm{EuTi}\textrm{O}_{3}$. Recently,
it was shown \citep{Sagarna2013} that doping $\textrm{EuTi}\textrm{O}_{3}$
by $\textrm{N}$ favors the tilted $Pnma$ structure also at high
temperatures. By means of hybrid functional calculations of electronic
structure it was shown that the presence of impurities as well as
the tilting of oxygen octahedra lead to delocalization of the Eu f-states,
which modifies f-d coupling between A and B sites and, therefore,
changes the lattice stability conditions. Note, that $Pnma$ is subgroup
of the $Imma$ considered in Ref. \cite{Rushchanskii2012}, and is
the one with off-centered Eu position. Thus, this particular field
is still open to research and further theoretical and experimental
studies are yet to come which will clarify the microscopic origin
of the observed Eu shift in EuTiO$_{3}$. 

The lattice dynamics in $\textrm{EuTi\ensuremath{\textrm{O}_{3}}}$
resembles that of $\textrm{SrTi\ensuremath{\textrm{O}_{3}}}$, the
atomic delocalization in phase change materials \citep{Matsunaga2011}
as well as the lattice dynamics in $\textrm{PbTe}$ \citep{Kirsten2012,Bozin2010}.
In such cases the associated potential energy is considered as a multi-valley
surface with drastical impact on thermal conductivity \citep{Christensen2010}.
The observed phenomenon resembles rattling between minima in the potential
energy and could be harnessed to lower the lattice thermal conductivity.

\section{Conclusion}

In summary, the combination of the extracted speed of sound by RUS
and NIS, the Gram - Charlier expansion of Eu atomic displacements
based on neutron diffraction and the feedback from theoretical studies
based on $ab\textrm{ }initio$ calculations provides evidence for
europium delocalization which originate in a lattice instabilities
in the system. Short range coexistence of crystallographic phases
with candidate symmetries\emph{ }$Imma$, \emph{$R\bar{3}m$} and
\emph{$I4/mcm$} in $\textrm{EuTi\ensuremath{\textrm{O}_{3}}}$ might
be related to the fact that Eu delocalization does not lead to a structural
transition. Experimental studies under high pressure on phase pure
EuTiO$_{3}$ single crystals or thermal diffuse scattering experiments
using synchrotron radiation might shed further light on this scenario.
In addition, according to theoretical calculations a Eu atomic displacement
is allowed in the $Imma$ phase and its calculated size \citep{Rushchanskii2012},
$0.012\textrm{ Å}$, corresponds roughly to the one measured in our
experiment, however, phase transition with long range order is not
observed. \textcolor{black}{Reasonable scenarios might be nucleation
of nanoclusters with }$Imma$ symmetry and different distortion direction
which fail to extend to a reasonable corellation length or dynamical
fluctuation of the distortion. Measurements of acoustic emission as
it was presented \citep{Dulkin2010} on BaTiO$_{3}$ might clarify
this scenario. Similar atomic off-centering has been observed in a
series of ferroelectric compounds such as $\textrm{Pb(Zr,Ti)}\textrm{O}_{3}$
and $\textrm{Pb}(\textrm{M}\textrm{g}_{1/3}\textrm{N}\textrm{b}_{2/3})\textrm{O}_{3}$
\cite{Egami2007}.

\section{Acknowledgements}

The Helmholtz Association of German research centers is acknowledged
for funding (VH NG-407 ``Lattice dynamics in emerging functional
materials'' and VH NG-409 \textquotedbl{}Computational Nanoferronics
Laboratory\textquotedbl{}). The European Synchrotron Radiation Facility,
the Advanced Photon Source, the PETRAIII and the Spallation Neutron
Source are acknowledged for provision of synchrotron radiation and
neutron beam time at ID22N, 6-ID-D,\textcolor{black}{{} P02.1}, POWGEN
and NOMAD respectively. The work in the Czech Republic has been supported
by the Czech Science Foundation (Project No. P204/12/1163), MEYS (LD
11035 and LD 12026 - COST MP0904) and the ERDF (CEITEC - CZ.1.05/1.1.00/02.0068).
We are grateful to Dr. D. Robinson, Dr. M. Feygenson and Dr. J. Neuefeind
for help during data acquisition and Jülich Supercomputing Center
for support.

\bibliographystyle{apsrev} 

\bibliographystyle{EuTiO3}
\bibliography{EuTiO3_short,EuTiO3}

\begin{thebibliography}{69}
\bibitem[{DOS()}]{DOS}
\bibinfo{note}{The DOS program was modified for reconvoluting the extracted DPS
  with a Gaussian function with the same FWHM as the measured instrumental
  function}.
\bibitem[{151()}]{151Eu}
\bibinfo{note}{The natural abundance of $^{151}\textup{Eu}$ is 47.8\% and no
  further isotopic enrichment was needed}.
\end{thebibliography}

\end{document}